\documentclass[usenatbib,useAMS,referee]{mn2e}
\usepackage[english]{babel}
\usepackage{graphicx,epsfig}
\usepackage{color}
\usepackage{morefloats}
\usepackage{amsmath,amssymb}
\usepackage{lscape}
\usepackage{hyperref}
\usepackage{url}
\usepackage{fancyhdr}

\title[High Energy Emission from 3C\,454.3]{Clues on High Energy Emission Mechanism from Blazar 3C\,454.3 during 2015 August Flare}

\author[]{Zahir Shah$^{1}$\thanks{shahzahir4@gmail.com}, S. Sahayanathan$^{2}$\thanks{sunder@barc.gov.in}, 
Nijil Mankuzhiyil$^{2}$, Pankaj Kushwaha$^{3}$, \newauthor Ranjeev Misra$^{4}$ and  Naseer Iqbal$^{1}$ \\
$^{1}$ Department Of Physics, University of Kashmir, Srinagar-190006, India \\
$^{2}$ Astrophysical Sciences Division, Bhabha Atomic Research Centre, Mumbai-400085, India\\
$^{3}$ Department of Astronomy (IAG-USP), University of Sao Paulo, Sao Paulo 05508-900, Brazil\\
$^{4}$ Inter-University Center for Astronomy and Astrophysics, Pune-411007, India}
\begin{document}
\date{}
\pagerange{\pageref{firstpage}--\pageref{lastpage}} \pubyear{2016}
\maketitle
\label{firstpage}
\begin{abstract}
We perform a detailed spectral study of a recent flaring activity from the Flat Spectrum Radio Quasar (FSRQ), 3C\,454.3, observed 
simultaneously in optical, UV, X-ray and $\gamma$-ray energies during 16 to 28 August, 2015. The 
source reached its peak $\gamma$-ray flux of $(1.9\pm0.2)\times\,10^{-05} \; {\rm ph\,cm^{-2}\,s^{-1}}$ 
on 22 August. The time averaged broadband spectral energy distribution (SED) 
is obtained for three time periods, namely 
``flaring state"; covering the peak $\gamma$-ray flux, ``post flaring state";
immediately following the peak flare and ``quiescent state"; separated from the flaring event and following the post flaring state.
The SED corresponding to the flaring state is investigated using different emission models involving synchrotron, synchrotron self Compton (SSC)
and external Compton (EC) mechanisms. Our study suggests that the X-ray and $\gamma$-ray emission from 3C\,454.3
cannot be attributed to a single emission mechanism and instead, one needs to consider both SSC and EC mechanisms. Moreover, the target photon energy responsible for the
EC process corresponds to an equivalent temperature of 564 K, suggesting that the flare location lies beyond the 
broad line emitting region of the FSRQ.  SED fitting of the other two flux states further supports these
inferences.
\end{abstract}

\section{Introduction} 

Blazars are the radio loud active galactic nuclei (AGNs), with a relativistic jet oriented close to the line of sight 
of the observer \citep{Urry1995}. They are characterized by a strong continuum emission extending 
from  radio  to  the $\gamma$-ray energies,  high  polarization at optical and radio frequencies, and a rapid flux
variability \citep{Sambruna2000,Fan2008}. Flat Spectrum Radio Quasars (FSRQs) are the subclass of blazars which, in 
addition, display broad emission lines and often a thermal ``blue bump" in the optical-UV region \citep{Urry1995}. 
Their luminosity is substantially higher than BL Lac objects, which are  another subclass of blazars \citep{Fossati98}. The spectral energy distribution (SED) 
of FSRQs are characterized by two prominent peaks, which 
are generally located at optical frequencies and hard X-ray/$\gamma$-ray regime. The  first  component  is  commonly attributed  
to   the synchrotron emission due to the interaction of relativistic distribution of electrons with the jet magnetic field; 
while the second peak is explained as inverse Compton (IC) scattering 
of low energy photons \citep{Urry1982, Begelman1987, Ghisellini1985, Blandford1995,Bloom1996,Sokolov2004}. 
If the low energy photons that undergo the IC scattering are the synchrotron photons, the process is known as Synchrotron Self Compton (SSC); whereas, if the low energy photons are from sources that are external to jet, then the IC process is termed as External Compton (EC).

The external photon field responsible for the EC emission could be either from the dusty torus \citep{Blazejowski2000}, and/or from the broad line region (BLR; \citealp{Dermer1993}). Identification of this external target photon field can hint about the location of the emission region in the blazar jet.
\cite{Sahayanathan2012} attributed the $\gamma$-ray emission from the FSRQ 3C\,279 to IC scattering of IR photons from dusty torus. The observed Very High Energy (VHE) spectra of this source is harder than the spectra which were predicted by Klein-Nishina process. 
Since, the EC scattering of BLR photons fall in Klein-Nishina regime, it was suggested that the EC/IR process as the 
favourable emission mechanism. A similar conclusion was arrived by \cite{Kushwaha2014} for the FSRQ PKS 1222+216. 
Alternatively, \cite{Cao2013} identified the location of the emission region of 16 FSRQs (of a sample of 21 FSRQs), 
which are beyond the BLR region, using the shape of their  $\gamma$-ray spectra.

3C\,454.3 is an FSRQ located at a redshift $z=0.86$ with coordinates (J2000) RA = 22:53:57.7 and DEC=16:08:54 \citep{Jackson1991}. 
It was first detected at $\gamma$-ray frequencies by EGRET on board \emph{CGRO} 
during a major outburst in 1992 \citep{Hartman1993}, in which the photon flux at energy E$>$100\,MeV was observed to vary 
from  $0.4 \times 10^{-6}\;\rm{ph}\,\rm{cm}^{-2}\rm{s}^{-1}$ to  $1.4 \times 10^{-6}\;\rm{ph}\,\rm{cm}^{-2} \rm{s}^{-1}$. In 2005, 
it underwent a major optical outburst reaching its historical peak with R band magnitude 12 \citep{Villata2006}. The flare was also 
observed at radio and X-ray frequencies, providing a simultaneous broadband SED \citep{Giommi2006}. However, the lack of 
$\gamma$-ray study during this period prevented further scrutinizing of the high energy emission mechanisms. During July and 
November of 2007,  3C\,454.3  was  active  again and the optical emission reached the level, that is comparable  to  that of  
2005 \citep{Raiteri2008}.  The AGILE satellite, detected intense $\gamma$-ray emission during these phases \citep{Vercellone2009,Vercellone2010}. 
The estimated flux during these periods were  $2.8\times 10^{-6}\;\rm {ph}\,{cm}^{-2} \rm{s}^{-1}$ and  
$1.7 \times 10^{-6}\;\rm{ph}\,{cm}^{-2} \rm{s}^{-1}$ respectively. The source was regularly monitored after the advent of Large Area 
Telescope (LAT) on board \emph{Fermi}, which witnessed many outbursts that were simultaneously observed at other frequencies. The 
flare detected in December 2009 was extremely luminous, where the flux at energy E$>$100\,MeV reached 
$\sim 2\times 10^{-5}\;\rm ph\,{cm}^{-2}{s}^{-1}$ \citep{Pacciani2010,Ackermann2010}.  In April 2010, it again became the 
brightest source in $\gamma$-rays with a flux level of  $\sim 1.5\times 10^{-5}\;\rm{cm}^{-2}\rm{s}^{-1}$ \citep{Ackermann2010}. 
November 2010 witnessed the most luminous $\gamma$-ray outburst from the source, which is a factor of 3 times brighter 
than the flaring state in December 2009, reaching up to a flux level 
of $\sim 6.8\times 10^{-5}\;\rm{ph}\,\rm{cm}^{-2}\rm{s}^{-1}$\citep{Striani2010,Vercellone2011,Abdo2011}. 
Though not comparable to the flare in 2010, the source again went on another high state in June 2014 to a flux level of  
$\sim 1.9\times 10^{-5}\;\rm{ph}\,{cm}^{-2}\rm{s}^{-1}$ \citep{Buson2014}. 

 In this work, we present a detailed spectral analysis of the broadband SED of 3C\,454.3, in order to explore the plausible high energy 
emission mechanisms. We analyse the \emph{Fermi} data corresponding to the $\gamma$-ray flare period  during August 2015, in order to obtain the spectra corresponding
to three time periods around the flare maximum. The $\gamma$-ray data are supplemented with simultaneous observations in X-ray/UV by
\emph{Swift}-XRT/UVOT and in optical by SMARTS. The broadband SEDs during this time period are studied in detail 
under simple emission models involving synchrotron, SSC and EC mechanisms. The parameter space is scrutinized thoroughly,
in the light of observed broadband spectral properties, to identify the plausible dominant emission mechanisms active at these observed
energy bands. Particularly, we study the possibility of associating X-ray and $\gamma$-ray emission under single IC 
emission process, and show that this interpretation fails to explain the observed properties, or demands unphysical parameters. 
Hence, we model the broadband SED, considering the synchrotron, SSC, and EC processes.
In the next section, we present the details of the observations and the data analysis procedure. In \S \ref{sec:model}, we describe the emission model and the approximate analytical formalism that we considered for the present work, and in \S \ref{sec:he} we apply this model on the SED of 3C\,454.3 to understand the high energy emission mechanism. Throughout the work, 
we use a cosmology with $\Omega_M = 0.3$, $\Omega_\Lambda = 0.7$ and $H_0$=71 km s$^{-1}$ Mpc$^{-1}$.

\section{Observation and Analysis}
\subsection{\emph{Fermi}-LAT}
\emph{Fermi} Large Area Telescope (LAT) is a pair conversion telescope which covers the  energy band from 20 MeV  to more than 300 GeV \citep{Atwood2009}. The data sample used for this analysis was obtained in the period of 16 to 28 August 2015.
The analysis was carried out using standard unbinned likelihood method \citep{Mattox1996} incorporated in the pylikelihood library of \emph{Fermi} Science Tools `v10r0p5' and the instrument response function (IRF)  ``P8R2\_SOURCE\_V6".  The event selection was based on Pass\,8 reprocessed contains only the SOURCE class events tagged as ``$\rm evclass=128,evtype=3$" with energies between 100 MeV and 300 GeV. The minimum contamination from Earth limb $\gamma$-rays is attained by barring the photons arriving from the zenith angle  $>90^{\circ}$. The data presented in this paper  were accessed from a $15^{\circ}$ radius region of interest (ROI) centred at the source  location (RA,DEC=343.490616, 16.148211). All the sources from the 3FGL catalogue within 15 degree ROI and an addition of 10 degree annular radii around it were modeled. The Galactic diffuse emission model and isotropic background used were ``gll\_iem\_v06.fit" and  ``iso\_p8R2\_SOURCE\_V6\_v06.txt" respectively. The contributory python package, ``make3FGLxml.py"  is used to create XML model file.  A maximum likelihood (ML) test statistics TS = 2$\Delta$ log(L) was used  to determine the significance of $\gamma$-ray signal. From the input model all the sources with TS $<$ 9 were deleted.  The unbinned likelihood analysis were repeated till it converged. The converged best fit values were then used to derive the flux and SED.  
\subsection{\emph{Swift}-XRT}
\emph{Swift}, a multi-wavelength (MWL) satellite is equipped with three telescopes: the Burst Alert Telescope (BAT; 15-150 keV; \citealp{Barthelmy2005}), the X-ray telescope (XRT; 0.3-10 keV; \citealp{Burrows2005}) and the UV/Optical Telescope (UVOT; 180-600 nm; \citealp{Roming2005}).\\  
The X-ray data which was taken in photon-counting mode were processed with the XRTDAS V3.0.0 software package. Standard XRTPIPELINE (Version: 0.13.1) were used to create the level 2 cleaned event files with default setting following the directions in the \emph{Swift} X-ray data analysis threads. To avoid pile-up, an annular region centred at the source positions were used to extract the source and background light curve (LC), spectra and image. The source region is extracted within the radii of 5 arcsec to 65 arcsec,  while the background region is selected  within the radii of 130 arcsec to 230 arcsec. The choice of these radii is based on  xrtgrblc V1.8  task \citep{Stroh2013}. XRTMKARF and GRPPHA  tasks were used to generate the ancillary response file and to rebin the energy channels, to have at least 20 counts per bin respectively. Spectral fitting with a single power-law in the energy range 0.3 to 10.0 KeV is performed using the XSPEC  version (12.8.2) \citep{Arnaud1996}. 
The absorbed power-law gives best fit to the data with the density of neutral hydrogen column ($n_H$) fixed to the Galactic value 
of $6.6\times10^{20} \,\rm{atoms\,cm^{-2}}$ \citep{Kalberla2005}, while the normalization and the spectral index were set as free parameters.

\subsection {\emph{Swift}-UVOT}

The UVOT data which was taken in the photometric filters --  V (5468 \AA), B(4392 \AA),
U(3465 \AA), UVW1(2600 \AA), UVM2(2246 \AA) and UVM2(1928 \AA)
\citep{Poole2008}-- were analysed using the task \emph{uvotsource} with a circular source region of 5 arcsec centered at the
source position, and a background region of radius 40 arcsec were selected
from nearby source free region.  The observed fluxes were corrected
for Galactic extinction using $E(B-V)=0.108$ and $R_V=A_V/E(B-V)=3.1$
following \cite{Schlafly2011}. 

\subsection{SMARTS}

The photometric optical observations were carried out using the Small and Moderate Aperture Research Telescope System (SMARTS). 
The data is acquired from the publicly available data archive \footnote{http://www.astro.yale.edu/smarts/glast/form.html}. 
The  details of data  reduction, acquisition, and calibration can 
be found in   \cite{Bonning2012}. The optical (B, V, R) and  near Infra red (J, K) are corrected for 
extinction following \cite{Schlafly2011}. 
The magnitudes are converted to physical flux units using the zero point fluxes of \cite{Bessell1998}.

\section{Analysis results}

\begin{figure}
\centering
\includegraphics[width=0.7\textwidth,angle=0]{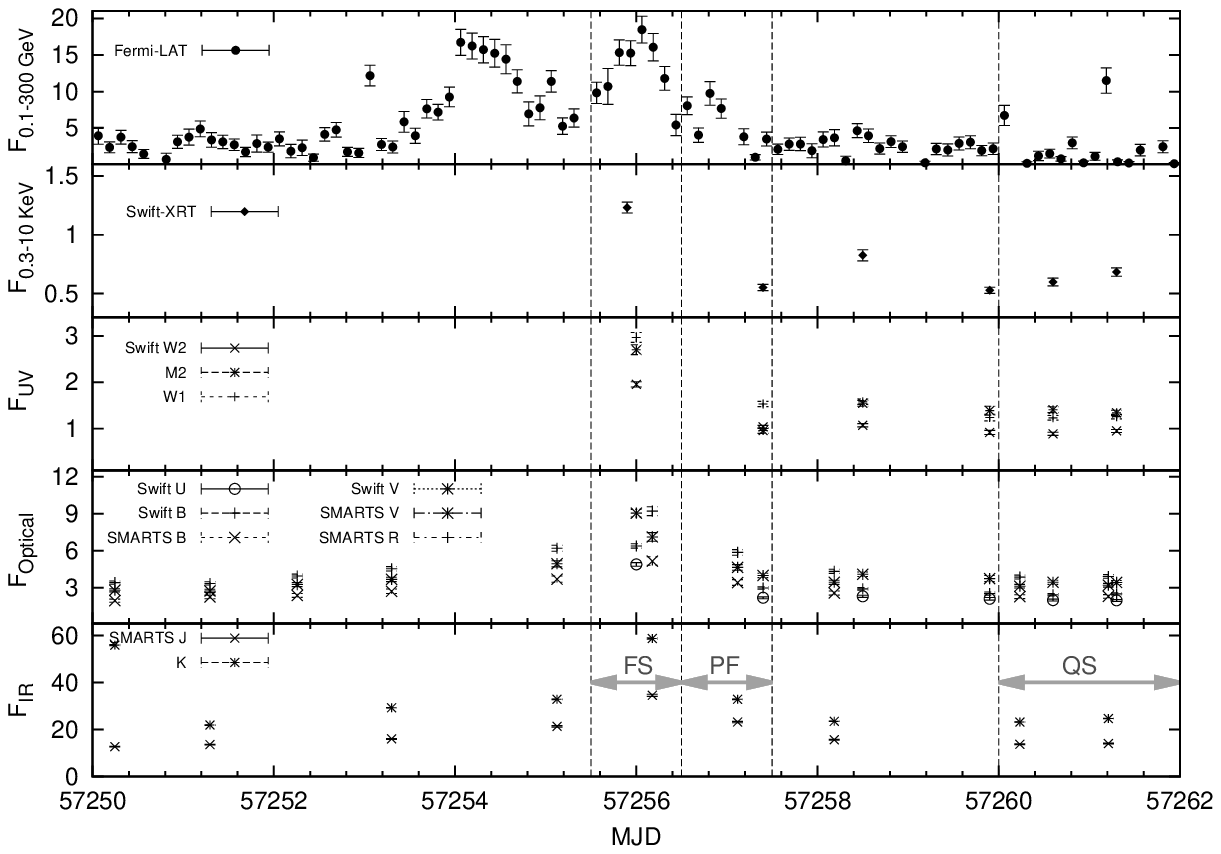}
\caption{ MWL LC of 3C\,454.3 during period MJD 57250-57262. $\gamma$-ray fluxes are in units of $10^{-6} \; {\rm ph\,cm^{-2}\,s^{-1}}$, X-ray fluxes are in units of $10^{-11}$ counts$\,s^{-1}$, and SMARTS fluxes are in units of mJy. Thick horizontal grey lines with arrow heads denote the time period for 
which time averaged SEDs corresponding to ``flaring state (FS)'', ``post flaring state (PF)'' and ``quiescent state (QS)'' are obtained.}
\label{fig:mwl}
\end{figure}

\begin{figure}
\centering
\includegraphics[width=0.5\textwidth,angle=270]{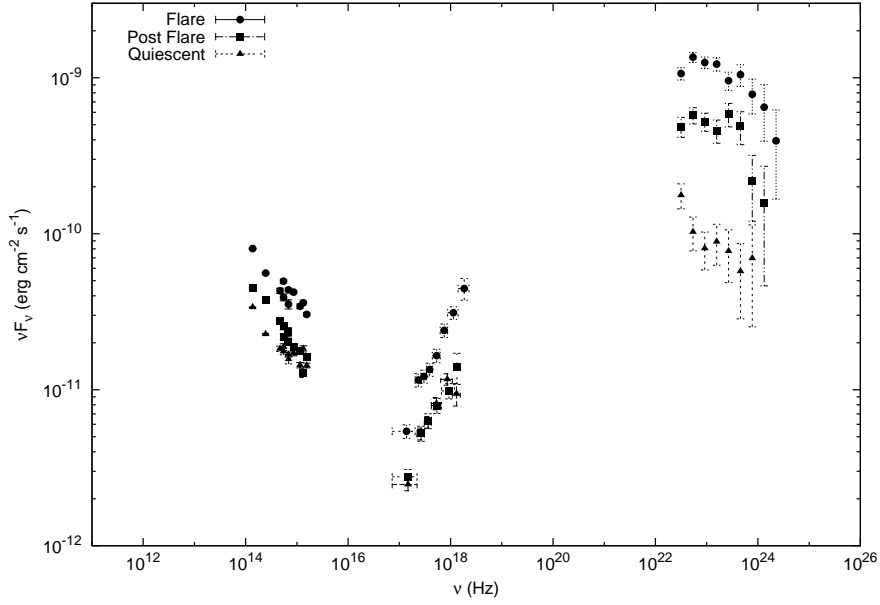}
\caption{Observed broad band SED of 3C\,454.3 obtained for three time periods, flaring state (MJD :57255.5-57256.5), post flaring state (MJD: 57256.5-57257.5) and quiescent state (MJD: 57260-57262) }
\label{fig:obsed}
\end{figure}

 Fig. (\ref{fig:mwl}) shows the obtained MWL LC of 3C\,454.3 during the period of 
 16 to 28 August 2015 (MJD: 57250-57262)  from \emph{Fermi}-LAT, 
 \emph{Swift}-XRT and \emph{Swift}-UVOT, and SMARTS observations. The good photon statistics in $\gamma$-ray frequency allow us to obtain finer LC down to 3\,hr binning which corresponds to a detection criteria of $TS > 9$ \citep{Mattox1996}.  Two peaks are clearly visible in the  $\gamma$-ray LC, where the first one peaks at MJD 57254.1 to a flux level of $(1.7\pm0.2) \times 10^{-5}\,{\rm ph\, cm^{-2} s^{-1}}$, while the second one peaks at MJD 57256.1 to a 
 flux level of $(1.9\pm0.2) \times 10^{-5}\,\rm{ph}\,\rm{cm^{-2}}\rm{s^{-1}}$.
 We restrict our study of flaring state on the second peak considering the better coverage of X-ray, UV and optical instruments during this period. To the best of our knowledge this is the third highest observed flux state of 3C\,454.3. By fitting the $\gamma$-ray flaring period of the LC with an exponential function, we found that the 
 shortest flux doubling time scale to be $t_{var,\,o}$ of $13.67 \pm 3.59$\,hr.
The X-ray LC peaks at  MJD 57255.9, corresponding to a flux 
of ($7.5 \pm 0.3) \times 10^{-11} \rm{erg}\,\rm{cm^{-2}}\rm{s^{-1}}$. 
After the flaring events, the flux decreases and the source returns to relatively low activity state. 
In order to study the spectral properties of this source in different flux states, we select three periods (which are well covered by $\gamma$-ray, X-ray, UV and optical instruments), 
namely flaring state (MJD 57255.5-57256.5), post flaring state (MJD 57256.5-57257.5) and quiescent state (MJD 57260-57262). These time periods are indicated by 
thick grey lines with arrow heads in Fig. (\ref{fig:mwl}).
The derived SEDs corresponding to these periods are shown in Fig. (\ref{fig:obsed}).  
The integrated fluxes and the spectral indices, obtained through a power-law fit of the $\gamma$-ray and X-ray data for each flux states are shown in Table\,\ref{table:lat} and Table\,\ref{table:xrt} respectively. The UV and optical fluxes during different states are summarized in  Table\,\ref{table:uvot} and Table\,\ref{table:smarts}. 
\par

\begin{table*}
\begin{tabular}{@{}lccc}
\hline
Flux state & Flux(0.1-300 GeV) & $\Gamma_{({\rm 0.1-300 GeV})}$ & TS\\
\hline
Flaring &  $14.80\pm0.56$  & $2.20\pm0.01$ & 3464.7 \\
Post Flaring &  $6.15\pm0.04$ & $2.15\pm0.05$ & 1179.6\\
Quiescent & $1.37\pm0.17$ & $2.40\pm0.12$ & 267.1  \\
\hline
\end{tabular}
\caption{Summary of power-law spectral fit to \emph{Fermi}-LAT observations. Col:- 1: Flux states, 
2: 0.1-300GeV integrated $\gamma$-ray flux in units of $10^{-6} {\rm ph\,cm^{-2}s^{-1}}$, 
3: Power-law indices and 4: Test statistics.}
\label{table:lat}
\end{table*}   

\begin{table*}
\begin{tabular}{@{}lccc}
\hline
Flux state & Flux(0.3-10 keV) & $\Gamma_{0.3-10 keV}$ & $\chi^2_{\rm red}$\\
\hline
Flaring & $6.71_{-0.53}^{+0.54}$  & $1.23_{-0.08}^{+0.09}$ & 1.1 \\ \\
Post Flaring & $2.74_{-0.3}^{+0.32}$ & $1.38_{-0.13}^{+0.12}$ & 0.6\\ \\ 
Quiescent & $2.7_{-0.23}^{+0.23}$ & $1.33_{-0.08}^{+0.09}$ & 1.4 \\ \\
\hline
\end{tabular}
\caption{Summary of power-law spectral fit to \emph{Swift}-XRT observations. Col.:- 1: Flux states,
2: 0.3-10 keV integrated flux in units of $10^{-11} {\rm erg\,cm^{-2}s^{-1}}$, 3: Power-law indices and
4: Reduced $\chi^2$ of fit statistics.}
\label{table:xrt}
\end{table*}

\begin{table*}
\begin{tabular}{@{}lcccccc}
\hline
Emission state & V & B & U & UVW1 & UVM2 & UVW2\\
\hline
Flaring  & $4.9\pm0.2$  & $4.3\pm0.11$ & $4.2\pm0.1$ & $3.4\pm0.1$ & $3.6\pm0.1$ & $3.0\pm0.1$ \\
Post Flaring  & $2.2\pm0.1$ & $2.0\pm0.1$ & $1.9\pm0.1$ & $1.8\pm0.1$ & $1.3\pm0.1$ & $1.6\pm0.1$\\
Quiescent   & $1.9\pm0.1$ & $1.7\pm0.1$ & $1.7\pm0.1$ & $1.4\pm0.1$ & $1.8\pm0.1$ & $1.4\pm0.1$\\
\hline
\end{tabular}
\caption{Summary of the \emph{Swift}-UVOT analysis. Col.:- 1: Flux states, 2-7: Flux at V, B, U, UVW1, UVM2 and UVW2 bands in units of $10^{-11}{\rm erg\; cm^{-2}s^{-1}}$.}
\label{table:uvot}
\end{table*}     
 
\begin{table*}
\begin{tabular}{@{}lccccc}
\hline
Emission state & B & V & R & J & K\\
\hline
Flaring & $3.5\pm0.2$  & $3.9\pm0.2$ & $4.3\pm0.2$ & $5.6\pm0.7$ & $8.0\pm0.4$ \\
Post Flaring & $2.3\pm0.2$ & $2.6\pm0.1$ & $2.7\pm0.1$ & $3.8\pm0.5$ & $4.5\pm0.2$\\
Quiescent & $1.6\pm0.1$ & $1.7\pm0.1$ & $1.8\pm0.1$ & $2.3\pm0.0$ & $3.3\pm0.0$\\
\hline
\end{tabular}
\caption{Summary of the SMARTS analysis. Col.:- 1: Flux states, 2-6: Flux at B, V, R, J and K bands in units of $10^{-11} {\rm erg\,cm^{-2}s^{-1}}$.}
\label{table:smarts}
\end{table*}

\section{Emission Model}\label{sec:model}
We model the optical/UV/X-ray/$\gamma$-ray emission from 3C\,454.3 using a simple one zone model 
involving synchrotron and IC emission processes. Under
this model the broadband emission from the source is assumed to arise from a spherical region 
of radius $R'$ moving down the jet with Lorentz factor $\Gamma$ at an
angle $\theta$ with respect to the observer.
The emission region is populated with a broken 
power-law distribution of electrons described by \footnote{All quantities with prime are measured in emission 
region frame, while with quantities subscript `o' are measured in the observer's frame (unless mentioned).}
\begin{align}
\label{eq:broken}
N'(\gamma') d\gamma' =\left\{
	\begin{array}{ll}
K \gamma'^{-p}d\gamma',&\mbox {~$\gamma'_{{\rm min}}<\gamma'<\gamma'_b$~} \\
K \gamma'^{q-p}_b \gamma'^{-q}d\gamma',&\mbox {~$\gamma'_b<\gamma'<\gamma'_{{\rm max}}$~}
\end{array}
\right.
\end{align}
where, $K$ is the normalization, $\gamma'$ is the electron Lorentz factor in the rest frame of the emitting region, $p$ and $q$ are the low and high energy indices 
of the electron energy distribution respectively, and $\gamma'_b$ is the Lorentz factor of the electron corresponding to the break in the distribution. 
The electrons lose their energy through synchrotron process in a tangled magnetic field $B'$ and by IC scattering
off the synchrotron photons and the ambient photons. Due to relativistic motion of the jet, the emission is boosted in the rest frame of AGN, determined by the Doppler factor 
$\delta = [\Gamma(1-\beta_\Gamma\, \rm{cos} \theta)]^{-1}$ where, $\beta_\Gamma$ is the velocity of the jet flow in units of speed of light, $c$. In case of aligned jets, e.g. blazars,
we can express $\delta=\zeta\Gamma$, with $\zeta \approx 2$ for $\theta = 0$ or $\zeta \approx 1$ for $\theta= \Gamma^{-1}$. 
If the flare duration is dominated by the light travel time, then the size of the emission region can be constrained from the flare 
time scale $t_{\rm var,\,o}$ as 
\begin{align} \label{eq:size_tvar}
	R'\lesssim \frac {\delta c\; t_{\rm var,\,o}}{1+z}
\end{align}
where, $z$ the redshift of the source. In addition, an energetically favoured configuration can be enforced by expressing
the magnetic field energy density $U_B'$ ($=B'^2/8\pi$) in terms of particle energy density $U_e'$, as $U_B' = \eta U'_e$. Here, $\eta\approx1$ corresponds to 
equipartition condition. The minimum energy of the emitting electron distribution; in case of shock acceleration, 
	can be assumed as $\Gamma$ \citep{Kino2002,Kino2004}. This is because, a thermal distribution peaking at energy  
	$\sim \Gamma$ is first formed by the shock, and from there, the electrons are accelerated further. Accordingly, we express
$\gamma'_{\rm min}= \chi \Gamma$ where, the factor $\chi$ decides the deviation from this scenario. Finally, after considering the relativistic and cosmological effects, the flux $F_{\rm {o}}$ received by 
the observer at frequency $\nu_o$, will be \citep{Begelman1984}
\begin{align}\label{eq:flux_obs}
F_{\rm o}(\nu_o)=\frac{\delta^3(1+z)}{d_{L}^2}V'\, j'\left(\frac{1+z}{\delta}\nu_o\right) \quad {\rm erg\,cm^{-2}s^{-1}Hz^{-1}}
\end{align}
Here, $d_{L}$ is the luminosity distance, $V'$ is the volume of emission region and $j'(\nu')$ is the emissivity at frequency $\nu'$ corresponding to different radiative 
processes (synchrotron, SSC and EC). 

An approximate analytical solution of the observed fluxes at a given frequency ($\nu_o$) can be obtained by expressing the single particle emissivities due to
synchrotron, SSC and EC processes as Dirac $\delta$-functions at their characteristic frequency (Appendix \ref{appendix:syn}, \ref{appendix:ssc}, \ref{appendix:ec} ). 
\begin{align}\label{eq:obs_syn_approx}
	{F}_{\rm o}^{\rm syn}(\nu_o)&\approx\left\{
\begin{array}{ll}
	\mathbb{S}(z,p)\,\delta^{\frac{p+5}{2}}B'^{\frac{p+1}{2}}
	R'^{3}K\nu_{o}^{-\left(\frac{p-1}{2}\right)}
	; \mbox {~$\nu_{o}\ll\nu_{p,\,o}^{\rm syn}$~} \\
\mathbb{S}(z,q)\,\delta^{\frac{q+5}{2}}B'^{\frac{q+1}{2}}
R'^{3}K\gamma_{b}'^{q-p}\nu_{o}^{-\left(\frac{q-1}{2}\right)}
; \mbox {~$\nu_{o}\gg\nu_{p,\,o}^{\rm syn}$~}
\end{array}
\right. \\
\nonumber \\
	\label{eq:obs_ssc_approx}
	{F}_{\rm o}^{\rm ssc}(\nu_o)&\approx \left\{
\begin{array}{ll}
	\mathbb{C}(z,p)\,\delta^{\frac{p+5}{2}}B'^{\frac{p+1}{2}}
	R'^{4}K^2\nu_o^{-\left(\frac{p-1}{2}\right)}\textrm{log}\left(\frac{\gamma'_b}{\gamma'_{\rm min}}\right)
	;\mbox {~$\nu_{o}\ll\nu_{p,\,o}^{\rm ssc}$~} \\
\mathbb{C}(z,q)\,\delta^{\frac{q+5}{2}}B'^{\frac{q+1}{2}}
R'^{4}K^2\gamma_b'^{2(q-p)}\nu_{o}^{-\left(\frac{q-1}{2}\right)} 
\textrm{log}\left(\frac{\gamma'_{\rm max}}{\gamma'_b}\right)
; \mbox {~$\nu_{o}\gg\nu_{p,\,o}^{\rm ssc}$~}
\end{array}
\right. \\
\nonumber \\
\label{eq:obs_ec_approx}
{F}_{\rm o}^{\rm ec}(\nu_{o})&\approx \left\{
\begin{array}{ll}
	\mathbb{E}(z,p)\,\delta^{p+3}U_*{\nu}_*^{\frac{p-3}{2}}
	R'^{3}K\nu_{o}^{-\left(\frac{p-1}{2}\right)}
	; \mbox {~$\nu_{o}\ll\nu_{p,\,o}^{\rm ec}$~} \\
	\mathbb{E}(z,q)\,\delta^{q+3}U_*{\nu}_*^{\frac{q-3}{2}}
	R'^{3}K\gamma_b'^{q-p}\nu_{o}^{-\left(\frac{q-1}{2}\right)}
	; \mbox {~$\nu_{o}\gg\nu_{p,\,o}^{\rm ec}$~}
\end{array}
\right.
\end{align}
Here, $\mathbb{S}$, $\mathbb{C}$ and $\mathbb{E}$ are functions of redshift and the particle 
index\footnote {The Equation (\ref{eq:obs_ssc_approx}) is not valid around $\nu_p^{\rm ssc}$, as we neglected the  cross scattering terms, where the synchrotron photons (emitted by the electrons with energy less than $\gamma'_b$) 
are scattered by electrons (with energy greater than $\gamma'_b$),  and vice versa.}. 
The external photon
field is assumed to be monochromatic with frequency $\nu_*$\footnote{Quantities with subscript `*' are measured in AGN frame} and energy density $U_*$. 
The observed synchrotron, SSC and EC 
peak frequencies are given by
\begin{align}
\label{eq:syn_peak}
\nu_{p,\,o}^{\rm syn}&=\frac{\delta}{1+z}\gamma_b'^{2}\nu_B \\
\label{eq:ssc_peak}
\nu_{p,\,o}^{\rm ssc}&=\frac{\delta}{1+z}\gamma_b'^4\nu_B \\
\label{eq:ec_peak}
\nu_{p,\,o}^{\rm ec}&=\frac{\delta}{1+z}\gamma_b'^2(\Gamma \nu_*)
\end{align}
where, $\nu_B=\frac{eB'}{2\pi m_ec}$ is the Larmor frequency.

\section{High Energy Emission from 3C\,454.3}\label{sec:he}

We apply the non-thermal emission model described in the previous section on the broadband SED of 3C\,454.3 available 
at optical, UV, X-ray and $\gamma$-ray energy bands (\S 2) to understand the dominance of different emission mechanisms
at these energies.
Since the low energy emission is well understood to be the synchrotron emission, we mainly concentrate on the high 
energy emission process. Before 
proceeding to the final model, we first test the different possible interpretations of the high energy emission. 

\subsection{ Considering only SSC}

Though the observed information -- photon spectral indices, optical/UV/X-ray/$\gamma$-ray fluxes, peak frequencies -- are not sufficient enough to constrain all the SSC parameters,
a relation between them can be obtained by solving the relevant approximate analytical solutions. Using synchrotron and SSC fluxes given by equations (\ref{eq:obs_syn_approx}) and  (\ref{eq:obs_ssc_approx}), together with the relations of
synchrotron and SSC peak frequencies given in equations (\ref{eq:syn_peak}) and  (\ref{eq:ssc_peak}), and by expressing $B'$ in terms of 
equipartition parameter $\eta$, one can obtain a relation between $\delta$ and the observable quantity $\nu_{p,\,o}^{\rm {syn}}$ as (Appendix \ref{appendix:sscdelta})
\begin{align}\label{eq:delta_syn_eta}
	\delta &\approx \, 1.4\times10^{-2}\,\left(\frac{\nu_{p,\,o}^{\rm ssc}}{3.5\times10^{22}\,\rm Hz}\right)^{-0.4}\left(\frac{F_{4.7\times10^{14}\, \rm Hz,\,o}^{\rm syn}}{9.2\times10^{-26}}\right)^{0.7}\left(\frac{F_{1.8\times10^{18}\, \rm Hz,\,o}^{\rm ssc} }{2.4\times10^{-29}}\right)^{-0.5}\nonumber\\
	&\times\,\left(\eta \, \mathbb{L}\right)^{-0.3}\,\left(\nu_{p,\,o}^{\rm syn}\right)^{0.1}  \, \left[\textrm{log}\left(\frac{\gamma'_b}{\gamma'_{\rm min}}\right)\right]^{0.5} 
\end{align}
Here, $F_{4.7\times10^{14}\,\rm{Hz},\,o}^{\rm syn}$ and $F_{1.8\times10^{18}\, \rm{Hz},\,o}^{\rm ssc}$ 
refer to the observed synchrotron and SSC fluxes at their corresponding mean frequencies 
in the observed optical and X-ray spectra, and $\mathbb{L}$ is given by equation (\ref{eq:L_gamma}). 
Particle indices $p=2.10$ and $q=4.18$ correspond to a photon spectral indices of $0.55$ and $1.59$ as 
observed in X-ray and $\gamma$-ray spectra respectively. The peak frequency of the high energy emission $\nu_{p,\,o}^{\rm ssc}=3.5\times 10^{22}$\,Hz is obtained by fitting a  
cubic polynomial to the observed X-ray and $\gamma$-ray fluxes shown in Fig. (\ref{fig:obsed}).

\begin{figure}
\centering
\includegraphics[width=0.5\textwidth,angle=270]{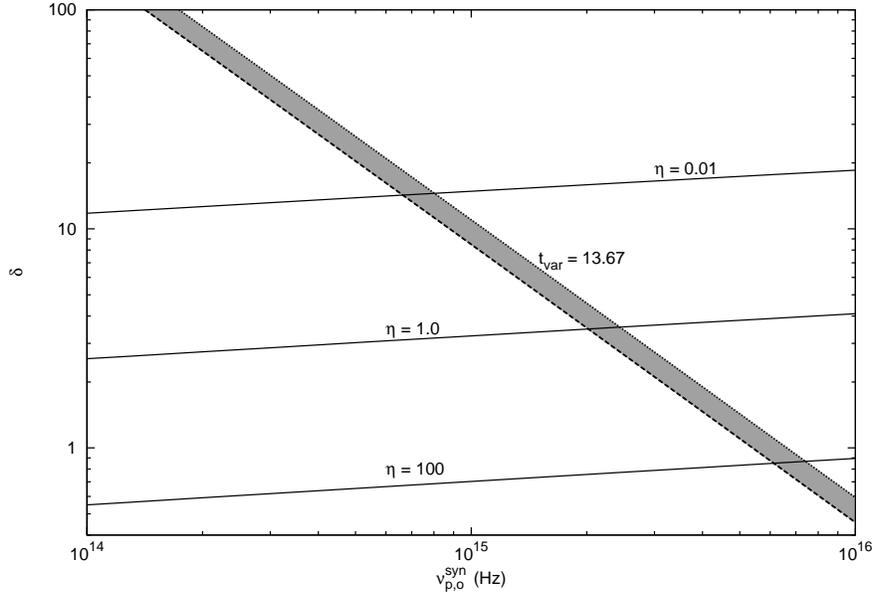}
\caption{Variation of $\delta$ with the $\nu_{p,o}^{\rm syn}$ for SSC process alone. The set of solid lines are from equation (\ref{eq:delta_syn_eta})  corresponds to $\eta=100,1.0,0.1$ and the 
grey band from equation (\ref{eq:delta_syn_tvar}) corresponds to variability time scale of $13.67 \pm 3.59$\,hr with lower bound denoted by dotted line and upper bound by dashed line. The observed $\nu_{p,o}^{\rm syn}\lesssim 10^{14}\,{\rm Hz}$, implies either a very small value of $\eta$ or a very long variability time scale.}
\label{fig:ssc}
\end{figure}

Alternatively, expressing $R'$ in terms of $t_{\rm var,\,o}$ and neglecting the equipartition condition between $U_B'$ and $U'_e$, one can
again obtain a relation between $\delta$ and $\nu_{p,\,o}^{\rm syn}$ as (Appendix \ref{appendix:sscdelta}).
\begin{align}\label{eq:delta_syn_tvar}
	\delta &\approx \, 2.1\times10^{34}\, \left(\frac{\nu_{p,\,o}^{\rm ssc}}{3.5\times10^{22} \rm Hz} \right)^{0.9}\,\left(\frac{F_{4.7\times10^{14}\rm  Hz,\,o}^{\rm syn}}{9.2\times10^{-26}}\right)^{0.5}\, \left(\frac{F_{1.8\times10^{18}\rm  Hz,\,o}^{\rm ssc}}{2.4\times10^{-29}}\right)^{-0.3} \nonumber\\ 	
	&\times \gamma_b'^{-1}\, t_{\rm var,\,o}^{-0.5} \left(\nu_{p,\,o}^{\rm syn}\right)^{-1.8}\,\left[\textrm{log}\left(\frac{\gamma'_b}{\gamma'_{\rm min}}\right)\right]^{0.3}  \quad
\end{align}
In Fig. (\ref{fig:ssc}), we show the correlation between $\delta$ and $\nu_{p,\,o}^{\rm syn}$ for the above two cases\footnote{Here and
	everywhere else, we choose $\chi\approx1$ and $\zeta\approx1$. In equations (\ref{eq:delta_syn_eta}) and (\ref{eq:delta_syn_tvar}), 
	$\gamma'_b$ and $\gamma'_{\rm min}$ can be expressed in terms of $\nu_{p,\,o}^{\rm syn}$ and $\delta$ respectively, and the equations are solved 
	iteratively to obtain the relation between $\nu_{p,\,o}^{\rm syn}$ and $\delta$.} with the choice of 
$\eta$ as $0.01$,$1$, and $100$, and the observed  variability time scale, $t_{\rm var,\,o}=13.67\pm3.59\, \rm{hr}$.
The observed optical spectrum suggests $\nu_{p,\,o}^{\rm syn}\lesssim 10^{14}\,{\rm Hz}$; which, either demands very long variability time scale $t_{\rm var,\,o} \gg 13.67$ hr, contrary to the observation, or too low
equipartition parameter $\eta\ll 0.01$ as shown in  Fig. (\ref{fig:ssc}). Based on this result, we conclude that SSC emission alone cannot explain the X-ray 
and $\gamma$-ray emission of the source successfully.

\subsection{Considering Only EC}
\begin{figure}
\centering
\includegraphics[width=0.5\textwidth,angle=270]{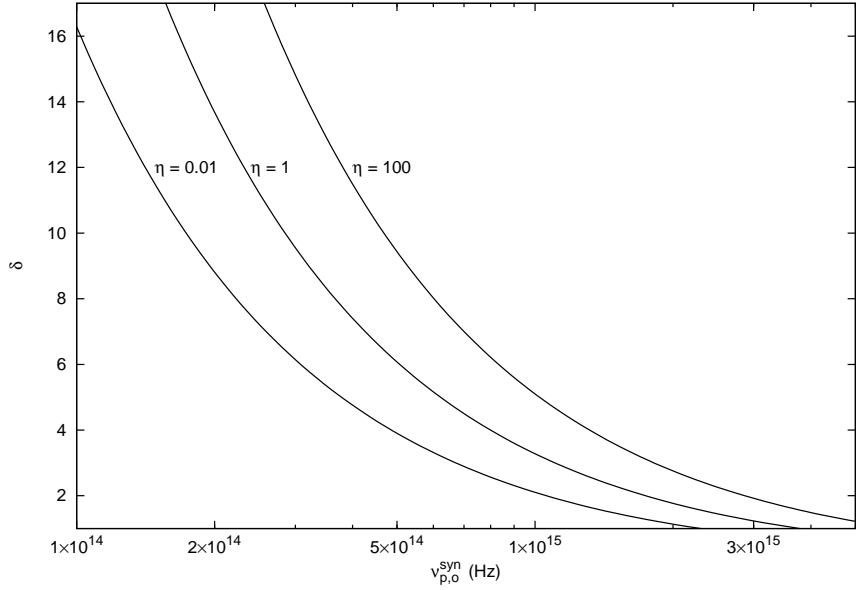}
\caption{Dependence of $\delta$ on $\nu_{p,o}^{\rm syn}$ described by equation (\ref{eq:delta_ec_eta}), for $\eta=0.01,1.0,100$ when the high energy emission is attributed to EC process alone. For $\nu_{p,o}^{\rm syn}\lesssim 10^{14}\,{\rm Hz}$, this demands $\delta>16$.}
\label{fig:eceta}
\end{figure}

Interpretation of the high energy emission from 3C\,454.3, as an outcome of EC process alone
requires two additional parameters -- equation (\ref{eq:obs_ec_approx}) --, namely monochromatic photon frequency ($\nu_*$) and energy density ($U_*$).  
If we consider the external photon field as a black body, the photon distribution will be significantly narrower than the non-thermal electron distribution. Hence, the energy density of the target photons can be estimated from the temperature which is obtained from the Wein displacement law.
\begin{align}\label{eq:bb}
	U_*=f\;\frac{4\sigma_{\rm SB}}{c}\left(\frac{h\nu_*}{2.82K_{B}}\right)^4
\end{align}
where, $f$ is the factor deciding the fraction of photons being IC scattered, $K_B$ is 
the Boltzmann constant and $\sigma_{\rm SB}$ is the Stefan-Boltzmann constant. Again, using approximate
synchrotron and EC fluxes equations (\ref{eq:obs_syn_approx}) and (\ref{eq:obs_ec_approx}), the peak
frequencies equations (\ref{eq:syn_peak}) and  (\ref{eq:ssc_peak}) and expressing $B'$ in terms of $\eta$,
and $R'$ in terms $t_{\rm var,\,o}$, the relation between $\delta$ and $\nu_{p,\,o}^{\rm syn}$ can be obtained as (Appendix \ref{appendix:ecdelta})
\begin{align}\label{eq:delta_ec_eta}
	\delta &\approx \, 10^{13}\,\left(\frac{\nu_{p,\,o}^{\rm ec}}{3.5\times10^{22}\rm Hz}\right)^{0.6}\, \left(\frac{F_{4.7\times10^{14} \rm Hz,\,o}^{\rm syn}}{9.2\times10^{-26}}\right)^{0.3}\, \left(\frac{F_{1.8\times10^{18}\rm Hz,\,o}^{\rm ec}}{2.4\times10^{-29}}\right)^{-0.2}  \nonumber\\
	&\times\left(\frac{t_{\rm var,\,o}}{13.67\, \rm hr}\right)^{-0.3}\,\eta^{0.1}\,\mathbb{L}^{0.1}\, \left(\nu_{p,\,o}^{\rm syn}\right)^{-0.9}
\end{align}
In Fig. (\ref{fig:eceta}), we show the plot between $\delta$ and $\nu_{p,\,o}^{\rm syn}$ for $\eta$=0.01, 1, 100 corresponding to  $t_{\rm var,\,o}=13.67$\,hr. 
Requirement of $\nu_{p,\,o}^{\rm syn}\lesssim 10^{14}$\,Hz based on the observed optical spectrum demands $\delta>16$.

\begin{figure}
\centering
\includegraphics[width=0.5\textwidth,angle=270]{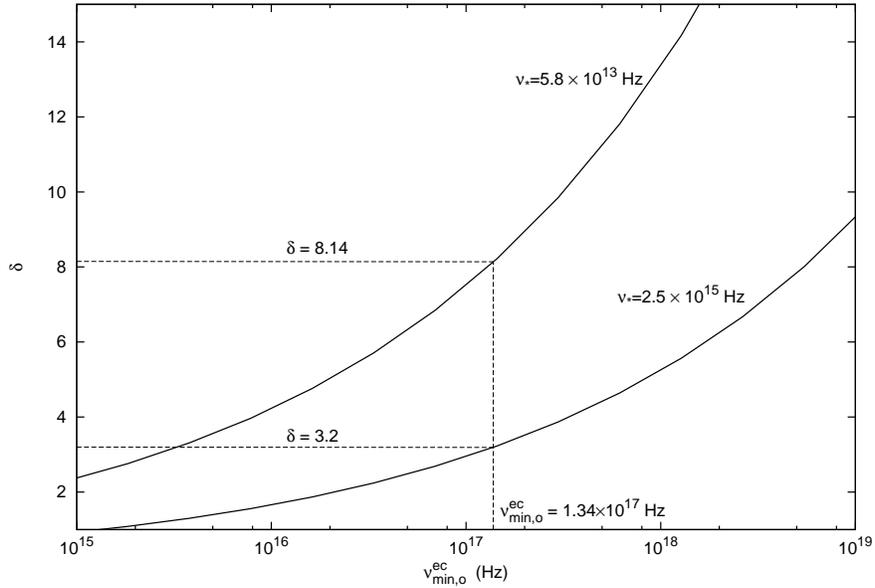}
\caption{Variation of $\delta$ with $\nu_{\rm min, o}^{\rm ec}$ described by equation (\ref{eq:ecmin}). The two solid curves correspond to target temperature 1000 K (IR torus) and 42,000 K (BLR). 
The vertical dashed line  is used to indicate minimum observed X-ray frequency during the flare and the horizontal dashed lines  represents  
$\delta$ corresponding to this frequency when the target photon temperature is 1000 K and 42,000 K respectively. Here $\delta=8.1$ and $\delta=3.2$, are in violation of the requirement $\delta>16$ obtained in Fig. (\ref{fig:eceta}).}
\label{fig:ecmin}
\end{figure}

The X-ray emission, under this interpretation, is produced by the IC scattering of external photons by low energy 
	electrons. The minimum frequency of the EC photon corresponds to the scattering of target photons by electrons with energy $\gamma'_{\rm min}$. Hence,
\begin{align}\label{eq:ecmin}
	\nu_{\rm min,\,o}^{\rm ec}&\approx \frac{\delta}{1+z}\gamma_{\rm min}'^2\Gamma\nu_*  \\
&\approx \frac{\chi^2}{\zeta^3} \frac{\nu_*}{1+z}\, \delta^4 \quad 
\end{align}
Here, $\Gamma \nu_*$ is the energy of the target photon measured in the emission region frame.
Considering the external target photon frequency that peaks either at $5.8\times10^{13}$ Hz (corresponds to the IR torus temperature
$\approx 1000$ K) or at $2.5\times10^{15}$  Hz (corresponds to the dominant Lyman alpha emission  from BLR), one can obtain the
relation between $\delta$ and $\nu_{\rm min,\,o}^{\rm ec}$ as shown in Fig. (\ref{fig:ecmin}). 
Clearly, the minimum observed X-ray frequency, $1.4\times10^{17}$ Hz, demands
$\delta\lesssim 8.1$ in case of IR emission and $\delta \lesssim 3.2$ for BLR photons. These values of $\delta$ are contradictory to the
one obtained earlier($\delta>16$) from equation (\ref{eq:delta_ec_eta}).
The minimum observed X-ray photon frequency and the condition $\delta>16$ can also be used to constrain $\gamma'_{\rm min}$ 
from equation (\ref{eq:ecmin}) as
\begin{align}\label{eq:ecmin1}
	\gamma'_{\rm min} < 3.2 \times 10^7 \sqrt{\frac{\zeta}{\nu_*}}
\end{align}
The external target photon field -- IR torus or BLR emission -- will then translate as 
$\gamma'_{\rm min} < 4$ or $\gamma'_{\rm min}<0.6$ respectively.
These values of $\gamma'_{\rm min}$ are either unphysical or much lower than the Bulk Lorentz factor of the jet.
Based on these studies, we assert that the X-ray emission and $\gamma$-ray emission 
may arise from different emission processes rather than a single emission process.

\subsection{Considering both SSC and EC}

\begin{figure}
\centering
\includegraphics[width=0.5\textwidth,angle=270]{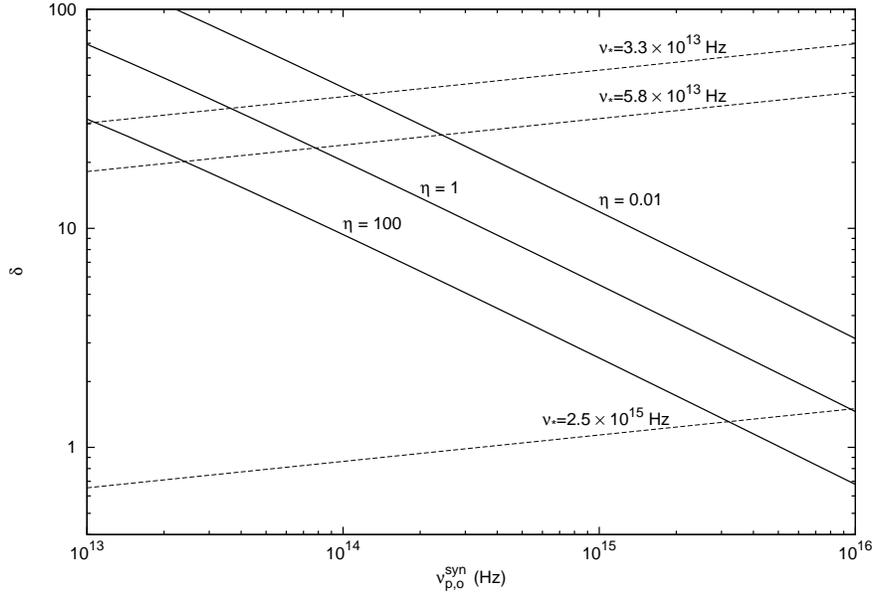}
\caption{Variation of $\delta$ with $\nu_{p,o}^{\rm syn}$. The solid lines correspond to the relation given by 
equation (\ref{eq:delta_ecssc_eta}) for the case of $\eta$ = 0.01, 1 and 100. The dashed lines correspond to the relation by equation (\ref{eq:delta_ecssc_t})
for the case of $T_*$ =  564 K, 1000 K and 42,000 K. For the latter, $f$ is fixed to 1.}
\label{fig:ecssc} 
\end{figure}

\begin{figure}
\centering
\includegraphics[width=0.5\textwidth,angle=270]{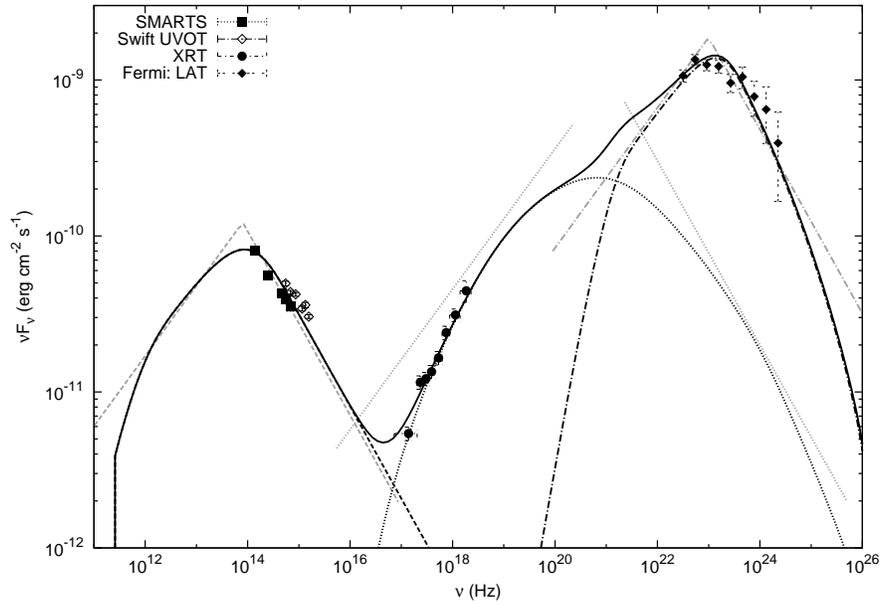}
\caption{SED of 3C\,454.3 during the flare state. The measured fluxes are shown in filled squares (SMARTS), open diamonds (\emph{Swift}-UVOT), filled circles (\emph{Swift}-XRT), and filled diamonds (\emph{Fermi}-LAT). Dashed line represents the 
synchrotron spectrum, while the dotted line and the dot-dashed line denote SSC spectrum and EC spectrum respectively. The solid line shows the total spectrum. Grey lines represent the analytical
approximation used, to estimated the model parameters.}
\label{fig:flaresed}
\end{figure}

\begin{figure}
\centering
\includegraphics[width=0.5\textwidth,angle=270]{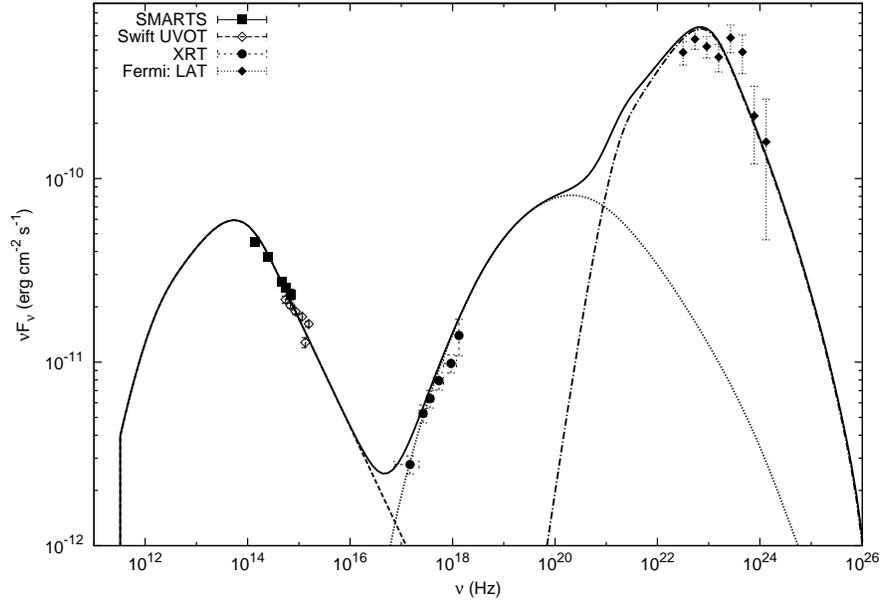}
 \caption{SED of 3C\,454.3 during the post flare state. The symbols used are the same as in Fig. (\ref{fig:flaresed}).}
\label{fig:pfsed}
\end{figure}

\begin{figure}
\centering
\includegraphics[width=0.5\textwidth,angle=270]{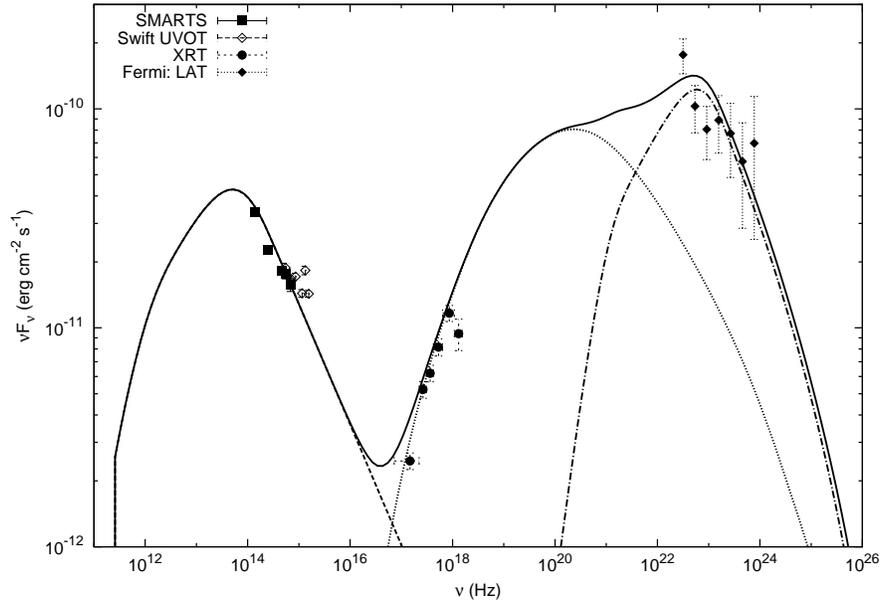}
\caption{SED of 3C\,454.3 during the quiescent state. The symbols used are the same as in Fig. (\ref{fig:flaresed}).}
\label{fig:qssed}
\end{figure}

Since a single emission mechanism is unable to explain the X-ray and $\gamma$-ray satisfactorily, we now explore the 
possibility of explaining the X-ray and $\gamma$-ray emission by SSC and EC processes and study the constraints
imposed on the underlying parameters. Attributing X-ray emission to SSC process, a relation between $\delta$ and $\nu_{p,\,o}^{\rm syn}$
can be obtained by solving equations (\ref{eq:obs_syn_approx}) and  (\ref{eq:obs_ssc_approx}) as (Appendix \ref{appendix:ecsscdelata})
\begin{align}\label{eq:delta_ecssc_eta}
	\delta &\approx 1.6\times10^{57} \, \left(\frac{F_{4.7\times10^{14} \rm Hz,\,o}^{\rm syn}}{9.2\times10^{-26}}\right)^{-1.7}\, \left(\frac{F_{1.8\times10^{18}\rm Hz,\,o}^{\rm ssc}}{2.4\times10^{-29}}\right)^{2.9}\, \left(\frac{t_{\rm var,\,o}}{17.3 \, \rm hr}\right)^{-6.4} \nonumber\\
	&\times \mathbb{L}^{4.1} \, \eta^{4.1}\, \gamma_b'^{19.9}\, \left[\textrm{log}\left(\frac{\gamma'_b}{\gamma'_{\rm min}}\right)\right]^{-2.9}\, (\nu_{p,\,o}^{\rm syn})^{-8.1}
\end{align}
Here, we have expressed $B'$ in terms of equipartition parameter $\eta$ 
and $R'$ in terms of $t_{\rm var,\,o}$. If $\gamma$-ray emission is dominated by EC process, again a relation between $\delta$ and $\nu_{p,\,o}^{\rm syn}$ can be 
obtained by solving equations (\ref{eq:obs_syn_approx}) and  (\ref{eq:obs_ec_approx}).
\begin{align}\label{eq:delta_ecssc_t}
	\delta &\approx 3.7\times10^9 \left(\frac{F_{4.7\times10^{14} \rm Hz,\,\,o}^{\rm syn}}{9.2\times10^{-26}}\right)^{-0.2}\, \left(\frac{F_{1.6\times10^{23} \rm Hz,\,o}^{\rm ec}}{7.8\times10^{-33}}\right)^{0.2}\, f^{-0.2}\, \gamma_b'^{-1}\, \nu_*^{-0.9}\, (\nu_{p,\,o}^{\rm syn})^{0.5}
\end{align}
In Fig. (\ref{fig:ecssc}), we plot the above two relations for different values of $\eta$, $\nu_*$ and $f$ \footnote{In equations (\ref{eq:delta_ecssc_eta}) and (\ref{eq:delta_ecssc_t}), 
$\gamma'_b$  can be expressed in terms of $\nu_{p,\,o}^{\rm syn}$ and $\delta$, $\gamma'_{\rm min}$ in terms of $\delta$, and the equations are solved 
iteratively to obtain the relation between $\nu_{p,\,o}^{\rm syn}$ and $\delta$.}. It can be noted that
the interpretation of the $\gamma$-ray emission by EC scattering of BLR photons demands $\nu_{p,\,o}^{\rm syn}\gg10^{14}$\,Hz, which is not supported
by the observation. However, the EC scattering of IR photons (EC/IR) at temperature $<1000$ K demands low values of 
$\nu_{p,\,o}^{\rm syn}$ which are within the observational constraints. Hence, we model the SED of 3C\,454.3 during different flux states
using synchrotron, SSC and EC/IR emission mechanisms. 
The parameters obtained from the approximate analytical expressions are further refined  in order to reproduce the observed fluxes using numerical emission models.  This  numerical model considers the exact single particle emissivity functions and Klein-Nishina corrected cross-section
for the IC processes (\citealp{Blumenthal1970}, \citealp {Dermer1993}). In Table \ref{table:sed}, we show the model parameters (top rows) corresponding
to these emission models for the three flux states chosen. In Fig. (\ref{fig:flaresed}), we show the model SED corresponding to flaring state, 
along with the observed fluxes.
The SED corresponding to approximate analytical expressions for synchrotron, SSC and EC processes (equations (\ref{eq:obs_syn_approx}),
 (\ref{eq:obs_ssc_approx}) and  (\ref{eq:obs_ec_approx}) are shown as grey lines. In Fig. (\ref{fig:pfsed}) and (\ref{fig:qssed}), we show the model SED
corresponding to post flare and quiescent states.   

\begin{table*}
\begin{tabular}{@{}lccc}
\hline
Parameters & Flaring & Post Flaring & Quiescent \\
\hline
$p$ & 2.1 &  2.1 & 2.1\\
$q$ & 4.18 &  4.18 & 4.1\\
$\gamma'_b$ & $2.5\times10^3$ & $1.7\times10^3$ & $1.8\times10^3$  \\
$U'_e$ & 0.1 & 0.04 & 0.06 \\
B' & 0.28 & 0.4 & 0.35 \\
$\eta$ & 25 & 7 & 12 \\
$\Gamma$ & 30 & 28.2 & 25.5 \\
$T_*$ & 564  & 658 & 583\\
f & 1 & 1 & 0.3\\ 
\hline
Properties &&&\\ 
\hline
$P_{\rm jet}$ & $3.5\times10^{46}$ &  $2.0\times10^{46}$ &  $2.1\times10^{46}$ \\
$P_{\rm rad}$ & $4.3\times10^{43}$ &  $2.4\times10^{43}$ & $1.3\times10^{43}$  \\
\hline
\end{tabular}
\caption{Source parameters and properties derived from the SED model at flaring, post flaring, and quiescent states. 
Row:- 1: Low energy power-law index of the particle distribution, 2: High energy power-law index of 
the particle distribution, 3: Electron Lorentz factor corresponding to the break energy of the particle distribution, 
4: Particle energy density in units of $\rm{erg\,cm^{-3}}$, 5: Magnetic field in units of $G$, 
6: Equipartition factor, 7: Bulk Lorentz factor of the jet flow, 8: Temperature of the external photon field in units of $K$, 9: Fraction of external photons participating
in the EC process, 10: Jet kinetic power derived from the source parameters assuming equal number of cold protons as of 
non-thermal electrons in units of  $\rm{erg\,s^{-1}}$ (see \S \ref{sec:dis}), 11: Total radiated power derived from the source parameters in units of  $\rm{erg\,s^{-1}}$ (see \S \ref{sec:dis}). 
The size of emission region is fixed at $R'=3.5\times10^{16}$ cm, viewing angle $\theta$ at $2^{o}$, $\gamma'_{\rm {min}}$ at 200 and $\gamma'_{\rm {max}}$ at $10^6$.}
\label{table:sed}
\end{table*}

\section{Discussion and Conclusion}\label{sec:dis}
The availability of simultaneous observation of 3C\,454.3 in optical, UV, X-ray and $\gamma$-ray, during August 2015 outburst
allow us to constrain the plausible emission mechanisms responsible for this emission. Detailed analysis of
various emission processes, during different flux states -- flaring, post-flaring and quiescent-- 
demands, the $\gamma$-ray emission to be dominated by EC process with
the target photon field in the IR regime. The obtained parameters during various flux states suggest that the flaring
behaviour of the source is mainly associated with the increase in the electron energy 
density and the bulk Lorentz factor (Table \ref{table:sed}). Since all the flux states have similar particle indices, together with minimum and maximum electron energies, the increase
in the electron energy density corresponds to the increase in the normalization constant $K$ of the electron energy
distribution. From equations (\ref{eq:obs_syn_approx}), (\ref{eq:obs_ssc_approx}) and (\ref{eq:obs_ec_approx}), 
the dependence of this quantity on the observed flux is linear, in the case of synchrotron and 
EC mechanisms; whereas, it is quadratic in the SSC mechanism. However, the increase in
flux during the flare in the $\gamma$-ray band is relatively larger than that of the optical band.  
This increase in flux is correlated with the increase in the bulk Lorentz factor,  which in turn enhances the target photon energy density
in the frame of the emission region by a factor $\Gamma^2$. We also note that the magnetic field of
the emission region and the break energy of the electron energy distribution are relatively less variant in the different flux states.

The estimated temperature of the target photon field can be used to constrain the location of the 
	emission region. For instance, the SED corresponding to the flaring state
suggests the IR target photon field to originate from a region with temperature $\approx 564$K. This region
can be associated with the outer part of the dusty torus covering the central AGN where such low temperatures
can be attained \citep{Jaffe2004}. If we consider this region to be illuminated by accretion disk, then the location of
this region can be estimated as  \citep{Pier1992}
\begin{align}
	D_{\rm IR} &\approx \frac{1}{T_*^2}\left(\frac{L_{\rm disk}}{4\pi\sigma}\right)^{\frac{1}{2}} \nonumber\\
	&\approx 3.8 \left(\frac{T_*}{564 K}\right)^{-2}\left(\frac{L_{\rm disk}}{10^{46}\rm erg \,s^{-1}}\right)^{\frac{1}{2}}\,\,\textrm{pc}
\end{align}
where $L_{\rm disk}$ is the accretion disk luminosity. Similarly, in the case of post flaring and quiescent states,
the location of the emission region can be estimated as $2.8$ pc and $3.5$ pc, which correspond to an 
external photon temperature of $\approx 658$ K and $\approx 583$ K respectively. Further, the EC scattering of IR photons to $\gamma$-ray 
energies also indicate that the most of the scattering process are confined within the Thomson regime, where the
scattering condition is
\begin{align}
\label{eq:thom_cond}
\gamma' \Gamma h\nu_*< m_e c^2
\end {align}
Here, $\Gamma \nu_*$ is the frequency of the target photon measured in the frame of the emission region. The frequency of the 
scattered photon will then be
\begin{align}
\label{eq:scatt_ph}
\nu_o' = \gamma'^2 \Gamma \nu_*
\end{align}
and in terms of observed frequency,
\begin{align}
	\label{eq:scatt_ph_obs}
	\nu_o \approx \frac{\zeta}{1+z} \gamma'^2 \Gamma^2 \nu_*
\end{align}
Using equations (\ref{eq:thom_cond}) and (\ref{eq:scatt_ph_obs}), we obtain
\begin{align}
	\nu_* &< \left(\frac{m_ec^2}{h}\right)^2 \left(\frac{\zeta}{1+z}\right)\nu_o^{-1} \nonumber \\
          &\lesssim10^{14} \left(\frac{\nu_o}{7\times10^{25}Hz}\right)^{-1} \left(\frac{\zeta}{1}\right) \quad {\rm Hz}
\end{align}
This frequency corresponds to an equivalent black body temperature of $\approx 1700$ K. This is larger than the temperature 
that we obtained (Table \ref{table:sed}), which validates  our approach of using Thomson condition for the derivation 
of physical parameters.

The kinetic energy of the jet can be estimated from the knowledge of the bulk Lorentz factor of the flow, provided
the mass density of the jet matter is known. The hadronic content of blazar jets are poorly understood and this 
prevents us from estimating this quantity. In leptonic models, similar to the
one considered here, hadrons are assumed to be cold and do not participate in the radiative processes. The absence
of hadronic signature in the SED, under such models, therefore hampers to put forth any reasonable estimation of the mass density
of the jet. However, an approximation of the same can be obtained by assuming the number of cold protons to be equal to 
that of non-thermal electrons. Under this approximation, the kinetic power of the jet can be estimated as \citep{Celotti1997}
\begin{align}
P_{\rm jet}=\pi R'^2 \Gamma^2 \beta_\Gamma c (U_B'+U'_e+U'_p)
\end{align} 
Here, $U_B'$, $U'_e$ and $U'_p$ are the co moving magnetic, leptonic, and hadronic energy densities. 
The total radiated power can be approximated from the emissivities corresponding to synchrotron, SSC and EC processes as
\begin{align}
	P_{\rm rad} \approx 4 \pi V'\int_0^\infty[j'_{\rm syn}(\nu')+ j'_{\rm ssc}(\nu') + j'_{\rm ec}(\nu')] d\nu'
\end{align} 
In Table \ref{table:sed} (bottom rows), we show the jet power and the total radiated power of
3C\,454.3 calculated during the three flux states considered here. 
For the chosen set of parameters, we find that, the radiated power at the blazar zone of the jet is much
less than the kinetic power. This ensures that only minimal amount of the jet power is expensed at the blazar zone
and most of the energy is retained to launch the jet up to kpc/Mpc scales.

The idea of using an additional IC component to explain the $\gamma$-ray emission in case of FSRQs 
was already perceived from the \emph{EGRET} 
(e.g. \citealp{Hartman01}) observations of blazars, and was later strengthened  by the \emph{Fermi} observations (e.g. \citealp{Abdo2010}). However, the choice of the
external target photon field was pre-assumed for  models advocating EC/BLR or EC/IR as a plausible mechanism (e.g. \citealp{Anderhub2009, Sikora2008}).
In the present work, we show that by exploiting the simultaneous information available at optical, X-ray and $\gamma$-ray 
energies, one can identify the target photon field using analytical approximation of different emissivity functions. Moreover,
 we also highlight the minimum necessary information (synchrotron flux, SSC flux, EC flux, peak frequencies) 
required to estimate the source parameters. In addition, we show that even in the absence of certain
information (e.g synchrotron peak frequency in this work, or the overshadowing of the synchrotron component by thermal emission) the source parameters can be inferred, by studying the allowed ranges of the missing information.

Alternate to the procedure presented in this work, the  GeV-TeV spectral shape of VHE detected FSRQs can also be used to identify the nature of EC mechanism. EC scattering of BLR photons  to VHE energies fall in Klein-Nishina regime, resulting a steep photon spectrum with index similar
to that of the emitting particle distribution \citep{Blumenthal1970}. On the other hand, EC scattering of IR
photons fall in Thomson regime with a relatively harder photon spectrum. This property of the EC
mechanisms is used to conclude the high energy emission process from \,3C\,279 as an outcome of EC/IR process \citep{Sahayanathan2012}. In the present work, we show that even for the sources where VHE
emission is not detected,
one can identify the high energy emission mechanism through broadband spectral 
modelling of optical--X-ray--$\gamma$-ray fluxes. Here, the EC scattering of BLR photons to low energy $\gamma$-rays
may still fall on Thomson region; hence, the emission mechanisms cannot be differentiated on the basis of the 
spectral slope. However, we exploit the spectral features of the broadband SED, like the peak frequencies, synchrotron 
and IC fluxes, to identify the most plausible emission mechanism, and use the same to constrain 
the target photon temperature.

The EC emission  of 3C\,454.3 during 2008 flare was explained using the BLR photons \citep{Anderhub2009}, while the break in the GeV spectra observed during August 2008, and November 2010 was  explained using EC scattering of both IR photons (from the dusty torus) and BLR photons (from the accretion disk), assuming a log parabola electron distribution assumption \citep{Cerruti2013}. However, the spectral features of the source during May-July 2014 suggest that the emission region is located close to the outer edge of the BLR region \citep{Britto2016}. Though we cannot assert, these differences, together with our results hint that, the photons responsible for EC emission at different periods are not always from the same region in the case of 3C 454.3.
This approach can be well extended to an ensemble of \emph{Fermi} detected FSRQs, for which simultaneous optical and X-ray information is available.  A detailed study to understand the emission parameters of broad sample of FSRQs will be addressed in a forthcoming work.

\section{Acknowledgement}
ZS thanks Atreyee Sinha, Vaidehi Paliya and Stalin C. S. for useful discussion and support in analysing the XRT and UVOT data. 
ZS, SS and NI are grateful to ISRO-RESPOND Program for the financial assistance under grant No. ISRO/RES/2/396. PK acknowledges support from FAPESP (2015/13933-0). This research has made use of data obtained from the High Energy Astrophysics Science Archive Research Center (HEASARC), provided 
by NASA Goddard Space Flight and optical data from SMARTS optical/near-infrared light. This work has made use of the XRT Data 
Analysis Software developed by the ASI Science Data Center (ASDC), Italy. We also thank an anonymous referee for providing constructive suggestions that have led to a substantial improvement of this manuscript. \\


\clearpage
\appendix
\section{Synchrotron flux coefficient ($\mathbb{S}$) }\label{appendix:syn}
The synchrotron emissivity due to a relativistic electron distribution $N'(\gamma')$ can be estimated using
\begin{align}
	j'_{\rm syn}(\nu')=\frac{1}{4\pi}\int\limits_1^\infty P_{\rm syn}(\gamma',\nu')N'(\gamma')d\gamma'
\end{align}
Here, $P_{\rm syn}(\gamma',\nu')$ is the single particle emissivity \citep{shu91}. Approximating $P_{\rm syn}(\gamma',\nu')$ as a $\delta$-function 
peaking at $\gamma'^2 \nu_B$, with $\nu_B = eB'/2\pi m_e c$ being the Larmor frequency, we obtain\citep{Sahayanathan2012} 
\begin{align}
j'_{\rm syn}(\nu')\approx \frac{c\,\sigma_T B'^2}{48\pi^2}\nu_B^{-\frac{3}{2}} N'\left(\sqrt{\frac{\nu'}{\nu_B}}\right)\nu'^{\frac{1}{2}}
\end{align}
Substituting this emissivity in the observed flux equation (\ref{eq:flux_obs}), we get
\begin{align}
	F_{\rm o}^{\rm syn}(\nu_o)\approx \frac{\delta^3(1+z)}{d_L^2}V' \frac{c\,\sigma_T B'^2}{48\pi^2}\nu_B^{-\frac{3}{2}}
	\sqrt{\frac{\nu_o(1+z)}{\delta}}\;N'\left(\sqrt{\frac{\nu_o(1+z)}{\delta \nu_B}}\right)
\end{align}
where, $\nu_o = \frac{\delta}{1+z}\gamma'^2 \nu_B$.
For broken power-law electron distribution given by equation (\ref{eq:broken}) we will obtain
\begin{align} \label{eq:app_syn}
	{F}_{\rm o}^{\rm syn}(\nu_o)&\approx\left\{
\begin{array}{ll} 	
	\mathbb{S}(z,p)\,\delta^{\frac{p+5}{2}}B'^{\frac{p+1}{2}}
	R'^{3}K\nu_{o}^{-\left(\frac{p-1}{2}\right)}
	&; \mbox {~$\nu_{o}\ll\nu_{p,o}^{syn}$~} \\
	\mathbb{S}(z,q)\,\delta^{\frac{q+5}{2}}B'^{\frac{q+1}{2}}
R'^{3}K\gamma_b'^{q-p}\nu_{o}^{-\left(\frac{q-1}{2}\right)}
&; \mbox {~$\nu_{o}\gg\nu_{p,o}^{syn}$~}
\end{array}
\right.
\end{align}
Here,
\begin{align}
\mathbb{S}(z,x)=\frac{c\,\sigma_T}{36\pi d_L^2}\left(\frac{e}{2\pi m_e c}\right)^{\frac{x-3}{2}}(1+z)^{\frac{3-x}{2}}
\end{align}
where, x can either be p or q.

\section{SSC flux coefficient ($\mathbb{C}$) }\label{appendix:ssc}
The IC emissivity due to a relativistic electron distribution $N'(\gamma')$ can be estimated using
\begin{align}
	j'_{\rm ssc}(\nu')=\frac{1}{4\pi}\int\limits_1^\infty P_{\rm ssc}(\gamma',\nu',\xi')N'(\gamma')d\gamma'
\end{align}
where, $P_{\rm ssc}(\gamma',\nu',\xi')$ is the single particle emissivity due to scattering of a target photon
of frequency $\xi'$. In case of SSC process, $\xi$ corresponds to the synchrotron photon and considering the
scattered photon frequency peaking at $\gamma'^2\xi'$, we obtain \citep{Sahayanathan2012}
\begin{align}
	j'_{\rm ssc}(\nu')\approx \frac{R'c}{36\pi^2}\sigma_T^2B'^2\nu_B^{-\frac{3}{2}}\nu'^{\frac{1}{2}}\int_{\gamma_{\rm min}}'^{\gamma_{\rm max}}\frac{d\gamma'}{\gamma'}N'\left(\frac{1}{\gamma'}\sqrt{\frac{\nu'}{\nu_B}}\right)N'(\gamma')
\end{align}
For a broken power-law electron distribution equation (\ref{eq:broken}), one can obtain
\begin{align}
	j'_{\rm ssc}(\nu')\approx \left\{
\begin{array}{ll}
	\frac{R'c}{36\pi^2}\sigma_T^2\,B'^2\nu_B^{-\frac{3}{2}}\nu'^{\frac{1}{2}}K^2 \textrm{log}\left(\frac{\gamma'_b}{\gamma'_{\rm min}}\right)\left(\frac{\nu'}{\nu_B}\right)^{-\frac{p}{2}}
	&;  \mbox {~$\nu'^{\rm ssc}_{min}<\nu'\ll\nu'^{\rm ssc}_{p}$~}\\
	\frac{R'c}{36\pi^2}\sigma_T^2B'^2\nu_B^{-\frac{3}{2}}\nu'^{\frac{1}{2}}K^2 \gamma_b'^{2(q-p)}\textrm{log}\left(\frac{\gamma'_{\rm max}}{\gamma'_b}\right)\left(\frac{\nu'}{\nu_B}\right)^{-\frac{q}{2}}
	&;  \mbox {~$\nu'^{\rm ssc}_{max}>\nu'\gg\nu'^{\rm ssc}_{p}$~} 
\end{array}
\right.
\end{align}
where, $\nu_{\rm min}'^{\rm ssc}=\gamma_{\rm min}'^4\nu_B$,\, $\nu_{\rm max}'^{\rm ssc}=\gamma_{\rm max}'^4\nu_B$ and $\nu_{p}'^{\rm ssc}=\gamma_{b}'^4\nu_B$.
Substituting this in flux equation (\ref{eq:flux_obs}), the observed SSC flux will be
\begin{align}\label{eq:app_ssc}
{F}_{\rm o}^{\rm ssc}(\nu_o)&\approx \left\{
\begin{array}{ll}
	\mathbb{C}(z,p)\,\delta^{\frac{p+5}{2}}B'^{\frac{p+1}{2}}
	R'^{4}K^2\nu_o^{-\left(\frac{p-1}{2}\right)}\textrm{log}\left(\frac{\gamma'_b}{\gamma'_{\rm min}}\right)
	&;\mbox {~$\nu_{o}\ll\nu_{p,o}^{ssc}$~} \\
\mathbb{C}(z,q)\,\delta^{\frac{q+5}{2}}B'^{\frac{q+1}{2}}
R'^{4}K^2\gamma_b'^{2(q-p)}\nu_{o}^{-\left(\frac{q-1}{2}\right)} 
\textrm{log}\left(\frac{\gamma'_{\rm max}}{\gamma'_b}\right)
&; \mbox {~$\nu_{o}\gg\nu_{p,o}^{ssc}$~}
\end{array}
\right. 
\end{align}
where,
\begin{align}
\mathbb{C}(z,x)=\frac{c\,\sigma_T^2}{27\pi d_L^2}\,\left(\frac{e}{2\pi m_e c}\right)^{\frac{x-3}{2}}\,(1+z)^{\frac{3-x}{2}}
\end{align}

\section{EC flux coefficient ($\mathbb{E}$) }\label{appendix:ec}
In case of EC process, the target photon is the Doppler boosted external photon field and for the case of  
a monochromatic photon field at frequency $\nu_*$ and energy density $U_*$, the EC emissivity can be approximated as \citep{Dermer1995} 
\begin{align}
	j'_{\rm ec}(\nu')\approx \frac{c\,\sigma_TU_*}{8\pi\nu_*}\sqrt{\frac{\delta\,\nu'}{\nu_*}}\;N'\left(\sqrt{\frac{\nu'}{\delta\,\nu_*}}\right)
\end{align}
The observed flux due to EC process can be obtained using equation (\ref{eq:flux_obs}) as
\begin{align}
	F_{\rm o}^{\rm ec}(\nu_o)\approx \frac{\delta^3(1+z)}{d_L^2}V' \frac{c\,\sigma_TU_*}{8\pi\nu_*}
	\sqrt{\frac{(1+z)\nu_o}{\nu_*}}\;N'\left(\frac{1}{\delta}\sqrt{\frac{(1+z)\nu_o}{\nu_*}}\right)
\end{align}
For broken power-law distribution of electrons equation (\ref{eq:broken}) we obtain
\begin{align} \label{eq:app_ec}
{F}_{\rm o}^{\rm ec}(\nu_{o})&\approx \left\{
\begin{array}{ll}
	\mathbb{E}(z,p)\,\delta^{p+3}U_*{\nu}_*^{\frac{p-3}{2}}
	R'^{3}K\nu_{o}^{-\left(\frac{p-1}{2}\right)}
	; \mbox {~$\nu_{o}\ll\nu_{p,o}^{\rm ec}$~} \\
	\mathbb{E}(z,q)\,\delta^{q+3}U_*{\nu}_*^{\frac{q-3}{2}}
	R'^{3}K\gamma_b'^{q-p}\nu_{o}^{-\left(\frac{q-1}{2}\right)}
	; \mbox {~$\nu_{o}\gg\nu_{p,o}^{\rm ec}$~}
\end{array}
\right.
\end{align}
where,
\begin{align}
\mathbb{E}(z,x)=\frac{c\,\sigma_T}{6\,d_L^2}(1+z)^{\frac{3-x}{2}}
\end{align}

\section{Estimation of Doppler factor from synchrotron and SSC processes}\label{appendix:sscdelta}
In order to interpret the broadband SED with respect to synchrotron and SSC process, we need information of 
five main free parameters namely, $B'$, $R'$, $K$, $\delta$ and $\gamma'_b$. These parameters can be 
expressed in terms of observed information using  the approximate expression for synchrotron flux and SSC flux given by 
equations  (\ref{eq:app_syn}) and  (\ref{eq:app_ssc}), the corresponding 
peak frequencies -- equations (\ref{eq:syn_peak}) and  (\ref{eq:ssc_peak}), and the equipartition condition $U_B' = \eta U'_e$.
A relation between the Doppler factor of the jet flow and the observable quantities can then be obtained as
\begin{align}\label{eq:delta_ssc_eta}
	\delta^{5-p} &\approx \frac{\mathbb{A}}{\mathbb{L}}\,  \frac{(F_{\rm o}^{\rm syn})^2}{\eta\,(F_{\rm o}^{\rm ssc})^{3/2}} \left(\frac{\chi}{\zeta}\right)^{p-2}(\nu_{p,o}^{\rm ssc})^{\frac{p-7}{4}} (\nu_{p,o}^{\rm syn})^{\frac{p-2q+7}{2}}\left[\textrm{log}\left(\frac{\gamma'_b}{\gamma'_{\rm min}}\right)\right]^{\frac{3}{2}}
\end{align}
where,
\begin{align}
	\mathbb{A}&=\frac{(p-2)}{8\pi m_e c^2} \frac{\mathbb{C}(z,p)^{\frac{3}{2}}}{\mathbb{S}(z,q)^2}\,\nu_{\rm syn,o}^{q-1}\nu_{\rm ssc,o}^{-\frac{3}{4}(p-1)}\left[\frac{2 \pi m_e c}{e}(1+z)\right]^{\frac{3p-4q+7}{4}}\\
        \label{eq:L_gamma}
		\mathbb{L}&=1-\left(\frac{\gamma'_{\rm min}}{\gamma'_b}\right)^{p-2}\, + \, \left(\frac{p-2}{q-2}\right)\left(\frac{\gamma'_{\rm min}}{\gamma'_b}\right)^{p-2}\left[1-\left(\frac{\gamma'_b}{\gamma'_{\rm max}}\right)^{q-2}\right]\\
		&\approx 1 \quad \textrm{for} \quad \gamma'_{min}\ll\gamma'_b\ll\gamma'_{max} \quad \& \quad q>p>2\nonumber
\end{align}
$F_{\rm o}^{\rm syn}$ and $F_{\rm o}^{\rm ssc}$ are the observed synchrotron and SSC fluxes at frequencies 
$\nu_{\rm syn,o}$($> \nu_{p,o}^{\rm syn}$) and $\nu_{\rm ssc,o}$($<\nu_{p,o}^{\rm ssc}$). Instead of using equipartition condition, an alternate relation between $\delta$ and the 
observable quantities can be obtained by  
expressing the emission region size $R'$ in terms of $t_{var,\,o}$ -- equation (\ref{eq:size_tvar}), as
\begin{align}
	\delta &\approx \mathbb{B}\,\gamma_b'^{\frac{p-q}{2}}\left(\frac{F_{\rm o}^{\rm syn}}{t_{\rm var,\,o}\sqrt{F_{\rm o}^{\rm ssc}}}\right)^{\frac{1}{2}}\left[\frac{\nu_{p,o}^{\rm ssc}}{(\nu_{p,o}^{\rm syn})^2}\right]^{\frac{2q-p+1}{8}} \left[\textrm{log}\left(\frac{\gamma'_b}{\gamma'_{\rm min}}\right)\right]^{\frac{1}{4}}
\end{align}
where,
\begin{align}
	\mathbb{B}=\frac{\mathbb{S}(z,q)^{\frac{1}{4}}}{\sqrt{\mathbb{C}(z,p)}}\left(\frac{1+z}{c}\right)^{\frac{5}{4}}\nu_{\rm ssc,o}^{\frac{p-1}{4}}\,\nu_{\rm syn,o}^{-\frac{q-1}{8}} \left[\frac{2\pi m_e c\,(1+z)}{e}\right]^{\frac{q-2p-1}{8}}
\end{align}
In the above equations, $\gamma'_b$ and $\gamma'_{min}$ can be replaced with 
\begin{align}\label{eq:gamma_b}
	\gamma'_b=\sqrt{\frac{\nu_{p,o}^{\rm ssc}}{\nu_{p,o}^{\rm syn}}}
\end{align}
and
\begin{align}
	\gamma'_{\rm min}=\frac{\chi}{\zeta}\,\delta
\end{align}

\section{Estimation of Doppler factor from synchrotron and EC processes}\label{appendix:ecdelta}
In case of synchrotron and EC processes, reproduction of SED requires 
knowledge of seven parameters, namely $B'$, $R'$, $K$, $\delta$, $\gamma'_b$, $U_*$ and $\nu_*$. 
These parameters can be expressed in terms of observed information using the approximate synchrotron 
and EC fluxes -- equations (\ref{eq:app_syn}) and  (\ref{eq:app_ec}), 
the corresponding peak frequencies -- equations (\ref{eq:syn_peak}) and  (\ref{eq:ssc_peak}), equipartition condition, 
relation between $R'$ and $t_{\rm var,\,o}$ -- equation (\ref{eq:size_tvar}) and assuming the target photon field as black-body 
-- equation (\ref{eq:bb}). A relation between $\delta$ and the observable quantities can then be obtained as 
\begin{align}
	\delta^{2p+6} &\approx \mathbb{D}\frac{\eta\, \mathbb{L}\,F_{\rm o}^{\rm syn}}{t_{\rm var,\,o}^3}\,\left[\frac{(\nu_{p,o}^{ec})^{(p+5)^2}}{(\nu_{p,o}^{syn})^{pq+9q+p+25}}\right]^{\frac{1}{8}}\left(\frac{F_{\rm o}^{\rm syn}}{F_{\rm o}^{\rm ec}}\right)^{\frac{p+5}{4}}\, \chi^{2-p}\, \zeta^{\frac{p^2+18p+9}{8}}
\end{align}
where,
\begin{align}
	\mathbb{D}&=\frac{8\pi m_e }{(p-2)c}\left(\frac{h}{2.82K_B}\right)^{p+5} \left(\frac{e}{2\pi m_e c} \right)^{\frac{pq+9q+p+25}{8}} (1+z)^{\frac{(p+9)(p-q)+24}{8}}\nonumber \\
	&\times \frac{1}{\mathbb{S}(z,q)}\left[\frac{4\,\mathbb{E}(z,p)\,f\sigma_{SB}}{\mathbb{S}(z,q)c}\right]^{\frac{p+5}{4} }\nu_{\rm syn,o}^{\frac{(q-1)(p+9)}{8}}\nu_{\rm ec,o}^{-\frac{(p-1)(p+5)}{8}}
\end{align}
$F_{\rm o}^{\rm ec}$ is the observed EC flux at frequency $\nu_{\rm ec,o}$ ($<\nu_{p,o}^{\rm ec}$).
Under this emission model, $\gamma'_b$ can be estimated as
\begin{align}
	\gamma'_b=\frac{1}{\delta}\sqrt{\frac{\zeta(1+z)\nu_{p,o}^{ec}}{\nu_*}}
\end{align}
and the target photon frequency is given by
\begin{align}
	\nu_*^{p+5}&=\mathbb{F} \left(\frac{\eta\mathbb{L}F_{\rm o}^{\rm syn}}{t_{var,\,o}^3}\right)^{2}\,\delta^{-4(p+3)}\, \zeta^{3p+1}\, (\nu_{p,o}^{\rm syn})^{-2(q+5)}\, (\nu_{p,o}^{\rm ec})^{p+5}
\end{align}
Here,
\begin{align}
	\mathbb{F}= \left[\frac{8\pi m_e}{c(p-2)\mathbb{S}(z,q)}\right]^2\left(\frac{e}{2 \pi m_e c}\right)^{q+5}\nu_{\rm syn,o}^{q-1}(1+z)^{p-q+6}
\end{align}

\section{Estimation of Doppler factor from synchrotron, SSC and EC processes}\label{appendix:ecsscdelata}
Reproduction of SED using synchrotron, SSC and EC processes again requires the 
knowledge of seven parameters, namely $B'$, $R'$, $K$, $\delta$, $\gamma'_b$, $U_*$ and $\nu_*$.
In this case, the SSC and EC peak frequencies cannot be estimated as they may fall anywhere between 
the X-ray and $\gamma$-ray band. However, we can obtain the information about the SSC and EC 
flux using the X-ray and $\gamma$-ray flux points. If we assume $\nu_*$ is known, corresponding to IR photons from 
dusty torus or the Lyman alpha emission from BLR, then one can obtain two relations between $\delta$ and 
the observable quantities. First, attributing X-ray emission to SSC emission, the equations/conditions 
required to express the parameters in terms of observed quantities can be obtained using the approximate synchrotron 
and SSC fluxes -- equations (\ref{eq:app_syn}) and  (\ref{eq:app_ssc}), synchrotron peak frequency -- 
equation (\ref{eq:syn_peak}), equipartition condition, and the 
relation between $R'$ and $t_{\rm var,\,o}$ -- equation (\ref{eq:size_tvar}). Here, we need only five equations/conditions
since interpretation using SSC does not require $U_*$. A relation between $\delta$ and 
the observable quantities can then be obtained as 
\begin{align}
	\delta^{p^2-2pq+2q+3} &\approx \mathbb{G}\,\frac{(F_{\rm o}^{\rm syn})^{p+1}}{(F_{\rm o}^{\rm ssc})^{q+1}}\left[\frac{\mathbb{L}\,\eta}{(\nu_{p,o}^{\rm syn})^2}\right]^{p-2q-1}t_{\rm var,\,o}^{4q-3p+1}
	\left(\frac{\chi}{\zeta}\right)^{-p^2+2pq+3p-4q-2}\nonumber \\
	&\times \gamma_b'^{p^2-pq+5p-9q-4}
	\left[\textrm{log}\left(\frac{\gamma'_b}{\gamma'_{\rm min}}\right)\right]^{q+1} 
\end{align}
Here,
\begin{align}
	\mathbb{G} &= \left[\frac{1+z}{\mathbb{S}(z,q)}\right]^{p+1}\mathbb{C}(z,p)^{q+1} \nu_{\rm syn,o}^{\frac{(q-1)(p+1)}{2}}\nu_{\rm ssc,o}^{-\frac{(p-1)(q+1)}{2}}c^{4q-3p+1}\left[\frac{\pi m_e(p-2)}{2e^2}\right]^{2q-p+1}
\end{align}
Then $\gamma'_b$ can be estimated from 
\begin{align}\label{eq:ecssc_gamma_b}
	\gamma_b'^{p^2-2p-19} &= \mathbb{H}\, (\mathbb{L}\eta)^4\, (\nu_{p,o}^{syn})^{p^2-2pq+2q+19}\,t_{\rm var,\,o}^{2(p-5)}
	\left[\textrm{log}\left(\frac{\gamma'_b}{\gamma'_{\rm min}}\right)\right]^{-({p+1})}  \nonumber \\
	&\times (F_{\rm o}^{\rm syn})^{2(1-p)} (F_{\rm o}^{\rm ssc})^{p+1}
\end{align}
and
\begin{align}
	\mathbb{H} &=\frac{\mathbb{S}(z,q)^{2(p-1)}}{\mathbb{C}(z,p)^{p+1}}\left(\frac{8\pi m_e c^2}{p-2}\right)^4 \left(\frac{c}{1+z}\right)^{2(p-5)}  \left[\frac{e}{2\pi m_e c (1+z)}\right]^{\frac{p^2-2pq+2q+19}{2}} \nonumber \\
&\times \nu_{\rm syn,o}^{-(q-1)(p-1)}\nu_{\rm ssc,o}^{\frac{(p-1)(p+1)}{2}}
\end{align}

The second relation between $\delta$ and the observable quantities can be obtained by using EC flux -- 
	equation (\ref{eq:app_ec}); instead of SSC flux, and expressing the external photon field as a black body 
	emission peaking at $\nu_*$ -- equation (\ref{eq:bb})
\begin{align}
	\delta^{q+1} &\approx \frac{\mathbb{I}}{f\, \nu_*^2} \frac{\bar{F}_{\rm o}^{\rm ec}}{F_{\rm o}^{\rm syn}} \left(\frac{\nu_{p,o}^{syn}}{\gamma_b'^2\,\nu_*}\right)^{\frac{q+1}{2}}
\end{align}
where,
\begin{align}
	\mathbb{I}=\frac{\mathbb{S}(z,q)}{\mathbb{E}(z,q)}\left[\frac{c}{4\sigma_T}\left(\frac{2.82K_B}{h}\right)^4\right] \left[\frac{2\pi\,m_e\, c}{e}(1+z)\right]^{\frac{q+1}{2}}\left(\frac{\bar{\nu}_{\rm ec,o}}{\nu_{\rm syn,o}}\right)^{\frac{q-1}{2}}
\end{align}
$\bar{F}_{\rm o}^{\rm ec}$ is the EC flux at frequency $\bar{\nu}_{\rm ec,o}$ ($>\nu_{p,o}^{\rm ec}$).

\label{lastpage}
\end{document}